\input harvmac
\input epsf

\noblackbox

\font\cmss=cmss10
\font\cmsss=cmss10 at 7pt

\newcount\figno
\figno=0
\def\fig#1#2#3{
\par\begingroup\parindent=0pt\leftskip=1cm\rightskip=1cm\parindent=0pt
\baselineskip=11pt
\global\advance\figno by 1
\midinsert
\epsfxsize=#3
\centerline{\epsfbox{#2}}
\vskip 12pt
{\bf Fig.\ \the\figno: } #1\par
\endinsert\endgroup\par
}
\def\figlabel#1{\xdef#1{\the\figno}}
\def\encadremath#1{\vbox{\hrule\hbox{\vrule\kern8pt\vbox{\kern8pt
\hbox{$\displaystyle #1$}\kern8pt}
\kern8pt\vrule}\hrule}}

\def\apm{\alpha'}
\def\l{\lambda}
\def\lpm{\lambda'}

\def\frac#1#2{{#1\over #2}}

\def\rlx{\relax\leavevmode}
\def\inbar{\vrule height1.5ex width.4pt depth0pt}
\def\IN{\relax{\rm I\kern-.18em N}}
\def\IP{\relax{\rm I\kern-.18em P}}
\def\ZZ{\rlx\leavevmode\ifmmode\mathchoice{\hbox{\cmss Z\kern-.4em Z}}
 {\hbox{\cmss Z\kern-.4em Z}}{\lower.9pt\hbox{\cmsss Z\kern-.36em Z}}
 {\lower1.2pt\hbox{\cmsss Z\kern-.36em Z}}\else{\cmss Z\kern-.4em
 Z}\fi}
\def\IZ{\relax\ifmmode\mathchoice
{\hbox{\cmss Z\kern-.4em Z}}{\hbox{\cmss Z\kern-.4em Z}}
{\lower.9pt\hbox{\cmsss Z\kern-.4em Z}}
{\lower1.2pt\hbox{\cmsss Z\kern-.4em Z}}\else{\cmss Z\kern-.4em
Z}\fi}
\def\IC{{\relax\,\hbox{$\inbar\kern-.3em{\rm C}$}}}

\def\narrowplus{\kern -.04truein + \kern -.03truein}
\def\narrowminus{- \kern -.04truein}
\def\narrowminussub{\kern -.02truein - \kern -.01truein}

\lref\bfss{T. Banks, W. Fischler, S.H. Shenker and  L. Susskind, 
``M Theory As A Matrix Model: A Conjecture,''
hep-th/9610043, Phys.Rev. { \bf D55} (1997) 5112-5128.}

\lref\jer{J.~Michelson,
``(Twisted) toroidal compactification of pp-waves,''
arXiv:hep-th/0203140.}

\lref\bmn{D.~Berenstein, J.~M.~Maldacena and H.~Nastase,
``Strings in flat space and pp waves from N = 4 super Yang Mills,''
JHEP {\bf 0204}, 013 (2002)
[arXiv:hep-th/0202021].}

\lref\gkp{S.~S.~Gubser, I.~R.~Klebanov and A.~M.~Polyakov,
``A semi-classical limit of the gauge/string correspondence,''
arXiv:hep-th/0204051.}

\lref\gmr{D.~J.~Gross, A.~Mikhailov and R.~Roiban,
``Operators with large R charge in N = 4 Yang-Mills theory,''
arXiv:hep-th/0205066.}
\lref\met{R.~R.~Metsaev,
``Type IIB Green-Schwarz superstring in plane wave Ramond-Ramond  
background,''
Nucl.\ Phys.\ B {\bf 625}, 70 (2002)
[arXiv:hep-th/0112044];
R.~R.~Metsaev and A.~A.~Tseytlin,
``Exactly solvable model of superstring in 
plane wave Ramond-Ramond  background,''
arXiv:hep-th/0202109.}
\lref\sv{M.~Spradlin and A.~Volovich,
``Superstring interactions in a pp-wave background,''
arXiv:hep-th/0204146.}

\lref\mand{S.~Mandelstam,
``Interacting String Picture Of The Neveu-Schwarz-Ramond Model,''
Nucl.\ Phys.\ B {\bf 69}, 77 (1974).}
\lref\gsb{M.~B.~Green, J.~H.~Schwarz and L.~Brink,
``Superfield Theory Of Type II Superstrings,''
Nucl.\ Phys.\ B {\bf 219}, 437 (1983).}
\lref\berk{N.~Berkovits,
``Conformal field theory for the superstring in a Ramond-Ramond plane  wave background,''
JHEP {\bf 0204}, 037 (2002)
[arXiv:hep-th/0203248].}
\lref\mrvv{S.Mukhi, M. Rangamani, E. Verlinde, H. Verlinde, work in progress.}
\lref\kpss{C.~Kristjansen, J.~Plefka, G.~W.~Semenoff and M.~Staudacher,
``A new double-scaling limit of N = 4 super 
Yang-Mills theory and PP-wave  strings,''
arXiv:hep-th/0205033.}
\lref\dm{S.~R.~Das and S.~D.~Mathur,
``Excitations of D-strings, Entropy and Duality,''
Phys.\ Lett.\ B {\bf 375}, 103 (1996)
[arXiv:hep-th/9601152].}
\lref\ms{J.~M.~Maldacena and L.~Susskind,
``D-branes and Fat Black Holes,''
Nucl.\ Phys.\ B {\bf 475}, 679 (1996)
[arXiv:hep-th/9604042].}
\lref\bn{D.~Berenstein and H.~Nastase,
``On lightcone string field theory from super Yang-Mills and holography,''
arXiv:hep-th/0205048.}

\lref\camb{N.~R.~Constable, D.~Z.~Freedman, M.~Headrick, 
S.~Minwalla, L.~Motl, A.~Postnikov and W.~Skiba,
``PP-wave string interactions from perturbative Yang-Mills theory,''
arXiv:hep-th/0205089.}
\lref\ggj{D. Ghoshal, R. Gopakumar, D. Jatkar, to appear.}
\lref\as{M.~Alishahiha and M.~M.~Sheikh-Jabbari,
``Strings in PP-waves and worldsheet deconstruction,''
arXiv:hep-th/0204174.}
\lref\motl{L.~Motl,
``Proposals on nonperturbative superstring interactions,''
arXiv:hep-th/9701025.}
\lref\dvv{R.~Dijkgraaf, E.~Verlinde and H.~Verlinde,
``Matrix string theory,''
Nucl.\ Phys.\ B {\bf 500}, 43 (1997)
[arXiv:hep-th/9703030].}

\lref\sen{A.~Sen,
``D0 branes on T(n) and matrix theory,''
Adv.\ Theor.\ Math.\ Phys.\  {\bf 2}, 51 (1998)
[arXiv:hep-th/9709220].}
\lref\seib{N.~Seiberg,
``Why is the matrix model correct?,''
Phys.\ Rev.\ Lett.\  {\bf 79}, 3577 (1997)
[arXiv:hep-th/9710009].}
\lref\mrv{S.~Mukhi, M.~Rangamani and E.~Verlinde,
``Strings from quivers, membranes from moose,''
JHEP {\bf 0205}, 023 (2002)
[arXiv:hep-th/0204147].}

\lref\myers{R.\ Myers, Dielectric-branes, [{\tt hep-th/9910053}].}

\lref\bs{T.~Banks and N.~Seiberg,
``Strings from matrices,''
Nucl.\ Phys.\ B {\bf 497}, 41 (1997)
[arXiv:hep-th/9702187].}

\Title
{\vbox{\baselineskip12pt
\hbox{hep-th/0205174}}}
{\vbox{\centerline{String Interactions in PP-Waves}}}

\centerline{Rajesh Gopakumar$^{a,b}$\foot{gopakumr@mri.ernet.in}}

\centerline{\sl $^a$ Harish-Chandra Research Institute, Chhatnag Rd.,}
\centerline{\sl Jhusi, Allahabad, India 211019.}
\centerline{\sl $^b$ Institute for Advanced Study, Olden Lane.,} 
\centerline{\sl Princeton, NJ08540, U.S.A.}
\medskip

\vskip 0.8cm

\centerline{\bf Abstract}
\medskip
\noindent

We argue that string interactions 
in a PP-wave spacetime are governed by an effective coupling
$g_{eff}=g_s(\mu p^+\apm)f(\mu p^+ \apm)$ where $f(\mu p^+ \apm)$ is
proportional to the light cone energy
of the string states involved in the interaction. 
This simply follows from  
generalities of a Matrix String description of this background. $g_{eff}$ 
nicely interpolates between the expected result ($g_s$) 
for flat space 
(small $\mu p^+\apm$ ) and a recently
conjectured expression from the perturbative gauge theory side 
(large $\mu p^+\apm$).

\vskip 0.5cm
\Date{May 2002}

\newsec{Introduction}

A recent important insight into the physics of large $N$,
${\cal N}=4$ Super Yang-Mills
theory has been that one might be able to isolate the dynamics of  
subsectors of the full theory, for instance, those 
carrying large global 
quantum numbers \bmn ,\gkp . 
Rather remarkably, BMN \bmn\ conjectured that a well defined subset of the
states in the gauge theory, carrying a large $U(1)$ R-charge, 
are really
the excitations of a Type IIB closed string in a PP-wave background geometry. 
\eqn\ppbkgd{
\eqalign{
ds^2 &= dx^-dx^+ - \mu^2\left(\sum_{i=1}^8 (x^i)^2\right)(dx^+)^2
+ \sum_{i=1}^8 (dx^i)^2\cr 
F^{(5)}_{+1234} &= F^{(5)}_{+5678} =\mu. }}
Moreover, free string theory in this background is exactly solvable in 
lightcone gauge \met . 
This allows one in principle to make direct comparisons 
between the gauge theory and the full string theory (beyond a supergravity
limit) \bmn\ . 

In particular, the spectrum of the first quantised strings is given in terms of
independent oscillators with the light cone energy of the $n$th oscillator 
being  
\eqn\en{
E_n=\mu\sqrt{1+{n^2\over (\mu p^+ \apm)^2}}.}
The total light cone energy of a generic string state is 
$E\equiv\mu f(\mu p^+ \apm)=\sum_n N_nE_n.$
The lightcone momentum $p^+$ is proportional to the $U(1)$ R-charge $J$ of 
the gauge theory 
$$
\mu p^+\apm= {J \over \sqrt{\l}}= {J\over g_{YM}\sqrt{N}}.
$$

The energy $E_n$ translates into the anomalous dimension in the gauge theory
of the corresponding operator 
$(\Delta -J)_n=\sqrt{1+{n^2\l \over J^2}}$.
For finite $g^2_{YM}=g_s$, the 'tHooft coupling blows up in the 
large $N$ limit. However, $(\Delta -J)_n$ is finite provided 
one scales $J\rightarrow \infty$ such that ${J^2\over N}$ is held fixed.
In fact, there are good indications that, rather than
$\l$, it is $\lpm={\l \over J^2}
={1\over (\mu p^+ \apm)^2}$ which is the right expansion 
parameter for perturbative gauge theory in this sector \bmn ,\gmr .

A formalism for studying string interactions in the lightcone 
framework
in this background has been developed  by 
\sv\ following the approach of \mand\ \gsb\ for flat space  
\foot{A covariant formalism has also been proposed in \berk .}. 
Interactions in a DLCQ framework are also being studied \mrvv .
On the gauge theory side, it has been observed by \kpss\ \bn\ \camb\ that
despite strictly taking $N\rightarrow \infty$, nonplanar diagrams survive,
being suppressed by factors of ${J^2 \over N}= g_s(\mu p^+ \apm)^2$. In other
words, contributions from 
gauge theory diagrams of genus $g$ are weighted with a factor of 
${J^{4g} \over N^{2g}}$. 
Thus it might be natural to guess that the three string interaction 
which governs splitting and joining of strings is weighted by an effective
string coupling of ${J^2 \over N}= g_s(\mu p^+ \apm)^2$. 

However, in this note we argue that the 
effective string coupling is actually 
$g_{eff}=g_s(\mu p^+ \apm)f(\mu p^+ \apm)$ where $E=\mu f(\mu p^+ \apm)$ 
is the total light cone energy  
involved in the interaction. 
In the most generic case involving a few ($O(1)$) oscillators 
with $O(1)$ excitations we see from Eq.\en\ that,
\eqn\genf{\eqalign{f(\mu p^+ \apm)\sim & \sqrt{1+{1\over (\mu p^+\apm)^2}}\cr 
\Rightarrow 
g_{eff}=& g_s(\mu p^+ \apm)\sqrt{1+{1\over (\mu p^+\apm)^2}}= 
{J^2\over N}\sqrt{{\l\over J^2}(1+{\l\over J^2})}}}
which is quite different from the above 
natural guess. 

Our reasoning 
is based on a second quantised Matrix String 
picture 
of interacting strings in the PP-wave background. The details of this Matrix 
String description will appear in a forthcoming publication \ggj . Here we will
obtain $g_{eff}$ simply by applying some scaling arguments that will
need only the general features of the Matrix String picture.

The $g_{eff}$ obtained here is in pleasing 
accord with a couple of facts. Consider the generic case 
where $f$ is as in 
Eq.\genf . Firstly, 
in the limit of small $\mu p^+\apm$ (or large $\lpm$) 
one recovers the flat space 
answer for $g_{eff}$, namely, simply $g_s$. 
Secondly, for large $\mu p^+\apm$ (or small
$\lpm$), $g_{eff}\sim g_s(\mu p^+\apm)$. In their
study from the perturbative gauge theory side, the authors of \camb\ 
were led to conjecture (for small $\lpm$) precisely
$g_s(\mu p^+\apm)={J^2\over N}\sqrt{\lpm}={J\over \sqrt{N}}g_{YM}$ 
as the effective interaction, at least amongst a class 
of string states. Thus our $g_{eff}$ provides the 
interpolating behaviour between these two regimes. 
We note in passing that our formula implies that for any small but
fixed $g_s$, (as well as fixed $(\mu p^+\apm)$) perturbation theory
breaks down if one scatters high energy string states.

\newsec{Matrix String Interactions}

\subsec{Flat Space}

We briefly recapitulate the Matrix String description of flat space.
 
Matrix String theory \motl\ \bs\ \dvv\ 
elegantly encapsulates the physics of weakly coupled second 
quantised strings in light cone gauge in terms of the IR dynamics of 
a $(1+1)$ dimensional gauge theory. For instance, in the case of Type IIA  
in $R^{10}$ with a null circle $x^-\sim x^-+2\pi R_0$ and 
$p^+={J\over R_0}$, the gauge theory is the maximally
supersymmetric $(1+1)$ dimensional $U(J)$ Yang-Mills theory
$$
S=\int d^2\sigma Tr_J\left(-{1\over 4}F_{\mu\nu}^2 +(D_{\mu}\Phi^i)^2 
+g_{YM}^2[\Phi^i,\Phi^j]^2  + fermions \right).
$$
The original parameters of the IIA lightcone string theory, namely, 
$R_0$ and the string coupling $g_s^{IIA}$ translate into the gauge theory 
parameters \seib\ 
$$g_{YM}={R_0\over g_s^{IIA}\apm}; ~~~~~~~~ \Sigma_1={\apm\over R_0}$$
where $\Sigma_1$ is the radius of the spatial direction in the Yang-Mills. 
  
The free string limit $g_s^{IIA}\rightarrow 0$ corresponds to the IR (strong
coupling) in the gauge theory. This is described by a free 
orbifold conformal field theory. 
The classical moduli space consists of 
commuting configurations $\Phi^j(\sigma)=diag\{{\phi^j_I(\sigma)}\},
~~ (I=1\cdots J)$.
Since the gauge symmetry includes the action by the Weyl group $S_J$, 
one can have
long string configurations $\Phi^j(\sigma+2\pi\Sigma_1)=g\Phi^j(\sigma)g^{-1}$,
where $g$ is an element of the Weyl group and hence permutes the eigenvalues.
The length of each permutation cycle in $g$ 
is the number of bits that go into 
forming a long string and is proportional to the $p^+$ it carries. The 
number of distinct cycles is the number of strings. In the limit 
$J\rightarrow \infty$ one can describe any number of free 
strings each carrying some 
arbitrary fraction of the total light cone momentum.

Similarly, for IIB string theory with $p^+={J\over R_0}$ \bs\
one first compactifies on a transverse 
circle of radius $R_1$, T-dualises to Type IIA,
lifts to M-theory and interchanges $x^-$ and $x^{11}$, (the 
``9-11 flip''). The end result is a theory of 
$J$ D-0 branes on two transverse circles in a decoupling limit \sen\ \seib . 
Or 
equivalently, the maximally supersymmetric $U(J)$ Yang-Mills in $(2+1)$ 
dimensions with the parameters \seib\
\eqn\param{
g_{YM}^2={R_0(R_1)^2\over g_s^{IIB}(\apm)^2},~~~~~~~~\Sigma_1={\apm\over R_0},
~~~~~~~~~~~\Sigma_2={\apm\over R_0}g_s^{IIB},}
where $\Sigma_i$ are the radii of the two spatial directions. 

The main difference from Type IIA 
arises in that even without going to weak coupling, if one
takes $R_1\rightarrow \infty$ (the noncompact limit of the original theory)
one is at strong Yang-Mills coupling. The resulting theory at the origin of 
the moduli space is the interacting 
maximally supersymmetric
$(2+1)$ dimensional CFT. The massless modes are parametrised by commuting
configurations 
$\Phi^j(\sigma_1,\sigma_2)=diag\{{\phi^j_I(\sigma_1,\sigma_2)}\},
~~ (I=1\cdots J)$. Again they can obey twisted boundary conditions but 
now in both spatial directions: 
$\Phi^j(\sigma_1+2\pi\Sigma_1, \sigma_2)=
g_1\Phi^j(\sigma_1,\sigma_2)g_1^{-1}$ and similarly with indices $1,2$ 
interchanged. To the extent that one can
classically describe this interacting theory, these configurations 
can be thought of as long membranes wrapped  
some number of times on both the cycles of the torus.

The simplification at weak coupling, $g_s^{IIB}\rightarrow 0$, is that 
$\Sigma_2\rightarrow 0$ and one has effectively a $(1+1)$ dimensional 
theory. The dimensional reduction is therefore 
the same orbifold CFT as
above, the only difference being that the fermions now have opposite 
chirality. The massless modes are thus long strings as in the IIA case, but 
now originate from the higher dimensional configurations
which are twisted only in the $\sigma_1$ direction.

\subsec{The PP-Wave Spacetime}
  
The matrix string description for the IIB PP-wave \ppbkgd\ 
closely follows the description for flat space \ggj .
We will study it as a limit 
$J,R_0\rightarrow \infty$ $(R_0=\mu\sqrt{\l}\apm)$ of a DLCQ 
theory with $p^+={J\over R_0}$ fixed (see \mrv\ \as\ for an explicit 
realisation of the finite $R_0$ geometry).

Despite appearances the background Eq.\ppbkgd\ has eight commuting
$U(1)$ isometries.  
One can compactify 
the Type IIB theory on one of these directions with radius $R_1$. 
T-duality takes one to a Type IIA background which can be 
lifted to M-theory \jer\
\eqn\mbkgd{
\eqalign{
ds_M^2 &= dx^-dx^+ - \mu^2\left(\sum_{I=3}^8 (x^I)^2+4(x^9)^2\right)(dx^+)^2
+ \sum_{i=1}^9 (dx^i)^2\cr 
F^{(4)}_{+129} &= -2\mu; ~~~~~F^{(4)}_{+349} =4\mu. }}
Reducing along $x^-$ gives a theory of $J$ D0-branes in a nontrivial 
curved background with 
two compact transverse directions $(x^1, x^2)$. The appropriate description
of the decoupled theory is 
again a $(2+1)$ dimensional $U(J)$ Yang-Mills 
theory with the same field content
as in the flat space case. But now with some 
number of additional pieces in the 
lagrangian
\eqn\extra{\eqalign{
\Delta S =& \int d^3\sigma (\Delta L_m +\Delta L_{myers}+\Delta L_{ferm});\cr
\Delta L_m \propto & \mu^2Tr_J(\Phi^i)^2; ~~~~~~~~~~\Delta L_{myers}
\propto \mu M_{ijk}Tr_J(\Phi^i\Phi^j\Phi^k).}}
It is easy to see the origin of these terms from Eq.\mbkgd . The mass
terms come from the nontrivial part of the metric, while the Myers terms
come from $F^{(4)}$.  
The corresponding mass and cubic terms 
involving fermions are determined by supersymmetry. 
This and other details, which will
appear in \ggj , will not be crucial for our scaling argument.

\subsec{String Interactions}

Let us momentarily
ignore the mass terms etc. in \extra . 
The weak coupling picture of strings splitting and joining emerges 
beautifully from the orbifold CFT description \dvv (which as we have seen is
the $g_s\rightarrow 0$ limit
of both IIA and IIB Matrix Strings, allowing for the 
differences in chirality). 
The leading irrelevant 
operator in the CFT is determined by the symmetries of the theory to be 
a dimension 3 operator, 
$V_{int}\propto {1\over M}\int d^2\sigma {\cal O}^{(3)}$, 
built out of the twist fields of the 
CFT. The twist fields lead to 
precisely the splitting and joining of strings by permuting eigenvalues.
A detailed study recovers the Mandelstam 
light-cone interaction vertex (delta function overlap) for these strings. 
We should think of this operator as arising, 
in the effective field theory language, from integrating out massive 
states whose dynamics we are not interested in. The mass parameter $M$
in $V_{int}$ 
is set by the lightest such state. In the Type IIB case, in reducing to the
$(1+1)$ dimensional theory, we have ignored KK excitations, the lightest of 
which has mass $M={1\over J\Sigma_2}$. This comes from 
a configuration carrying fractional momentum in the $\sigma_2$ direction \dm\
\ms .

Turning on the mass terms in \extra\ will not seriously
disturb this effective
field theory description provided 
we consider energies much less than $M$. 
To be precise, we want 
the quadratic part of the Hamiltonian (which we denote by 
$H_0$ and which includes the mass terms) to have energies much less than $M$.
The relevant configurations will then continue to be diagonal matrices
with the $S_J$ action. In other words, at the energy scales set by $H_0$, the 
states correspond to the second quantised Fock space of free strings in
the PP-wave spacetime. The Matrix String action in this limit
is essentially a (diagonal) matrix version of the Green-Schwarz action
in the PP-wave background (as is appropriate for a description of 
multiple strings).

Since the eigenvalues of $H_0$ are the light cone energies of the 
free string excitations involved in the interaction (Eq.\en\ and below), we 
need $E=\mu f(\mu p^+\apm)<<M$. 
Using Eq. \param\ 
this translates into 
$g_{eff}=g_s(\mu p^+\apm)f(\mu p^+\apm)<<1$ 
as the condition for the
DVV picture of string interactions to hold. 

Stated differently, from effective field theory reasoning,
the dimensionless coupling that suppresses the higher dimension operator
${\cal O}^{(3)}$
is really $g_{eff}={E\over M}$ and we can rewrite the above
interaction vertex as,   
$$V_{int}\propto {g_{eff}\over E}\int d^2\sigma {\cal O}^{(3)}.$$
As mentioned in the introduction, this effective coupling $g_{eff}$
interpolates between flat space behaviour and that suggested by 
perturbative gauge theory.

We should remark here that the operator ${\cal O}^{(3)}$
should be thought of as the DVV interaction operator but now dressed up by
RG flow to energy scales $E$. The mass deformation preserves
the original $S_J$ orbifold action.
This and the constraint of supersymmetry
should hopefully enable one to obtain an explicit expression for 
${\cal O}^{(3)}$ in terms of twisted sector 
modes of the free (but now massive) orbifold theory. This would
be important for using the Matrix String description to do 
explicit string computations. We hope to address this issue in the future.
It would also be very interesting to compare the results here with 
that from light cone string field theory as developed in \sv\ .
   
\newsec{ Acknowledgements}

I would like to thank my collaborators of \ggj , D. Ghoshal and D. Jatkar
for a lot of discussions on this topic.
I should also thank K. Hori, J. Maldacena, S. Mukhi, B. Pioline,
M. Rangamani, N. Seiberg, 
S. Sethi, M. Spradlin and S. Wadia for helpful conversations.
I must specially thank S. Minwalla for stimulating
discussions as well as for useful comments 
after a reading of the manuscript. The author's 
work has been generously supported by the people of India. He also
acknowledges support from the DOE grant DE-FG02-90ER40542. 

\listrefs

\end